# Band tail interface states and quantum capacitance in a monolayer molybdenum disulfide field-effect-transistor

**Nan Fang[1] and Kosuke Nagashio[1,2]**

[1]Department of Materials Engineering, The University of Tokyo, Tokyo 113-8656, Japan

[2]PRESTO, Japan Science and Technology Agency (JST), Japan

E-mail: nan@adam.t.u-tokyo.ac.jp, nagashio@material.t.u-tokyo.ac.jp

**Abstract.** Although MoS$_2$ field-effect transistors (FETs) with high-$k$ dielectrics are promising for electron device applications, the underlying physical origin of interface degradation remains largely unexplored. Here, we present a systematic analysis of the energy distribution of the interface state density ($D_{it}$) and the quantum capacitance ($C_Q$) in a dual-gate monolayer exfoliated MoS$_2$ FET. The $C_Q$ analysis enabled us to construct a $D_{it}$ extraction method as a function of $E_F$. A band tail distribution of $D_{it}$ with the lowest value of $8\times10^{11}$ cm$^{-2}$eV$^{-1}$ suggests that $D_{it}$ is not directly related to the sharp peak energy distribution of the S vacancy. Therefore, the Mo-S bond bending related to the strain at the interface or the surface roughness of the SiO$_2$/Si substrate might be the origin. It is also shown that ultra-thin 2D materials are more sensitive to interface disorder due to the reduced density of states. Since all the constituents for the measured capacitance are well understood, $I$-$V$ characteristics can be reproduced by utilizing the drift current model. As a result, one of the physical origins of the metal/insulator transition is suggested to be the external outcome of interface traps and quantum capacitance.

KEYWORDS: MoS$_2$, FET, quantum capacitance, interface states, density of states, MIT

## 1. Introduction

MoS$_2$ field-effect transistors (FETs) with high-$k$ dielectrics have attracted significant attention in ultimate scaled device research [1-7] because a natural thin body (0.65 nm per layer) is expected to suppress short-channel effects. The electrostatic field-effect control of carriers determines most of the device characteristics and needs to be fully understood before exploring the underlying physics of the electrical transport properties. For example, metal-insulator-transition (MIT) is widely studied in MoS$_2$ and other two-dimensional (2D) materials [3,8-10]. However, a poor understanding of high-$k$/MoS$_2$ interface properties might result in an erroneous subsequent physical analysis because the field-effect control by the gate is extrinsically affected by the interface states density ($D_{it}$), which may arise from the defects in MoS$_2$ and/or dangling bonds in high-$k$ oxides. Specifically, detailed observation via scanning transmission electron microscopy (STEM) and scanning tunneling microscopy (STM) has indicated the existence of sulfur vacancies on the order of ~$10^{13}$ cm$^{-2}$ in mechanically exfoliated and chemically vapor deposited (CVD) MoS$_2$, which introduce defect states below the conduction band according to the density functional theory (DFT) calculation [11-13] and severely degrade its electrical properties [14-17]. However, it has not yet been determined whether the electrically activated interface states originate from defect states corresponding to S vacancies because they were evaluated as a function of gate voltage instead of the Fermi energy ($E_F$).

Moreover, the field-effect control by the gate is reduced intrinsically due to the small density of states ($DOS$) of thin MoS$_2$ [18,19]. Extra kinetic energy is required to induce carriers in the MoS$_2$ channel, which introduces quantum capacitance ($C_Q = e^2 DOS$). The evolution of equivalent circuits from bulk MoS$_2$ to monolayer MoS$_2$ is shown in **figure S1**. The capacitance for the multilayer MoS$_2$ consists of both the depletion capacitance ($C_{Dep}$) and $C_Q$ in series, whereas $C_Q$ is the only constituent for monolayer MoS$_2$ because $C_{Dep}$ becomes so large that it can be neglected. Here, capacitance-voltage ($C$-$V$) measurement is powerful for directly probing both $C_Q$ and $D_{it}$, which results in the full understanding of the mechanism of field-effect control [20,21]. Although researchers have attempted to extract $D_{it}$ as a function of $E_F$ for multilayer and monolayer MoS$_2$ with both the capacitor structure [10,17,22,23] and FET [16,24-26] structure by $C$-$V$ measurements, the lack of a detailed study on $C_Q$ makes the $D_{it}$ energy distribution questionable. Based on the $DOS$ of 2D materials and the Fermi distribution, $C_Q$ is theoretically formulated for the MoS$_2$ monolayer [19], but it is neither measured well nor fitted experimentally [24]. One of the main reasons is the lack of consideration of how the interface trap capacitance ($C_{it} = e^2 D_{it}$) affects $C_Q$ extraction. Therefore, to elucidate all the constituents of the electrostatic field-effect control, focus should be on monolayer MoS$_2$, which finally results in the understanding of the whole picture of transport properties in the MoS$_2$ FET.

In this work, the systematic investigation of *C-V* and current-voltage (*I-V*) measurements of the same samples is carried out based on relatively high quality monolayer mechanically exfoliated MoS$_2$ FETs. The interface properties are evaluated as a function of $E_F$ to elucidate the physical origin of interface degradation. A band-tail-shaped $D_{it}$ is observed with the lowest value of $8\times10^{11}$ cm$^{-2}$eV$^{-1}$ for the monolayer MoS$_2$. With careful consideration of the effect of interface states, $C_Q$ is clearly extracted experimentally over the temperature range of 75-300 K for the first time. The correlation between top gate bias ($V_{TG}$) and $E_F$ is obtained via the $C_Q$ analysis. Having evaluated $C_{it}$ and $C_Q$ quantitatively by *C-V* measurements, *I-V* characteristics are then well reproduced and understood by utilizing the drift current model. The origin of MIT in monolayer MoS$_2$ is finally discussed and is suggested to be the external outcome resulting from $C_{it}$ and $C_Q$.

## 2. Results and discussion

In this paper, monolayer MoS$_2$ films were mechanically exfoliated onto the SiO$_2$ (90 nm)/$n^+$-Si substrate from natural bulk MoS$_2$ flakes. Raman spectroscopy and atomic force microscopy (AFM) were employed for determining the layer number (Details are shown in **figure S2**). Ni/Au was deposited as source/drain electrodes. Then, Y metal with a thickness of 1 nm was deposited via thermal evaporation of the Y metal from a PBN crucible in an Ar atmosphere with a partial pressure of 10$^{-1}$ Pa, followed by oxidization in the laboratory atmosphere at room temperature to form the buffer layer [27,28]. The Al$_2$O$_3$ oxide layer with a thickness of 10 nm was deposited via atomic layer deposition, followed by the Al top-gate electrode formation. *I-V* and *C-V* measurements were conducted using Keysight B1500 and 4980A LCR meters, respectively. All electrical measurements were performed in a vacuum prober with a cryogenic system.

### 2.1. $D_{it}$ extraction from S.S. in I-V

**Figure 1** shows a schematic drawing and optical image of the dual-gate monolayer MoS$_2$ FET. It should be noted that a monolayer MoS$_2$ with a large area (>30 μm$^2$) was selected for device fabrication and characterization because the measured capacitance should be larger than the stray capacitance (~10 fF) of the measurement system. **Figure 2a,b** shows the $I_{DS}$–$V_{TG}$ characteristic at $V_{DS}$ = 0.1 V as a function of $V_{TG}$ for three different MoS$_2$ FET samples measured at room temperature. The device performance is often scattered from device to device, indicating relatively low reliability of monolayer MoS$_2$. The top gate oxide capacitance ($C_{TG}$) can be determined by the relative ratio of capacitive coupling between the top and back gates with a MoS$_2$ channel (Details are provided in **figure S3** and the extracted physical properties are summarized in **table S1**). According to the $C_{TG}$ value, the two-probe field effect mobilities ($\mu_{FE}$) for samples 1, 2 and 3 are estimated to be 9.5, 6.0 and 2.0 cm$^2$ V$^{-1}$ s$^{-1}$, respectively. Although the mobility is largely underestimated due to the access region, as indicated in **figure 1b**, and the contact resistance, its difference among these three samples still indicates the difference in their interface properties. Indeed, the sample with the highest mobility (sample 1) exhibits the sharpest subthreshold region, in other words, the smallest subthreshold swing (*S.S.*). The *S.S.* values for the $I_{DS}$ range of 10$^{-11}$~10$^{-10}$ A for samples 1, 2 and 3 are estimated to be 110, 300, 300 mV/dec, respectively. Since *S.S.* depends on $V_{TG}$, $D_{it}$ can be precisely extracted as a function of $V_{TG}$-$V_{TH}$ in the subthreshold region based on *S.S.*, as shown in **figure 4d**, where $V_{TH}$ is the threshold voltage (Details are provided in **note S1**). The sample with the highest mobility has the lowest $D_{it}$ level within the smallest $V_{TG}$ range.

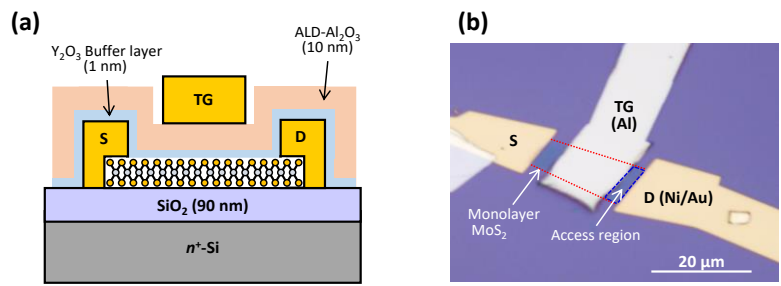

**Figure 1.** (a) Schematic diagram of the dual-gate monolayer MoS$_2$ FET. S, D and TG indicate the source, drain and top gate electrodes, respectively. (b) Optical image of the device. The access region refers to the channel region not covered by the top gate electrode.

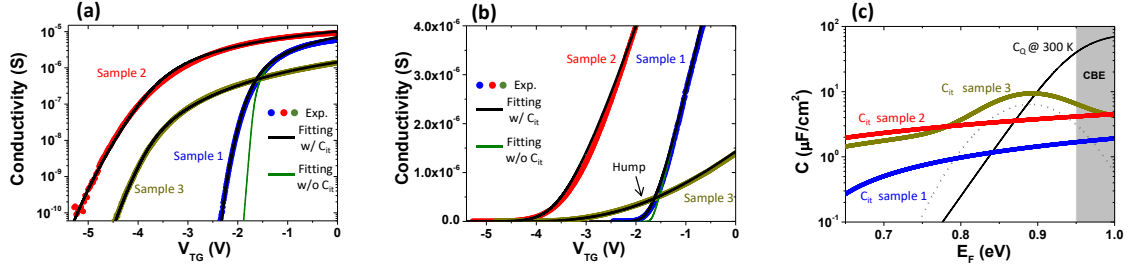

**Figure 2.** (a) Subthreshold transport characteristics of three different MoS$_2$ FET samples ($V_{DS}$ = 0.1 V). All data are measured at room temperature. No leakage current through the top gate is detected, which is the noise level (~pA) for the entire $V_{TG}$ range. (b) Linear scale of (a). (c) Three kinds of $C_{it}$ and $C_Q$ as a function of $E_F$ used for I-V modeling of samples 1-3. Band-tail-shaped $C_{it}$ are assumed for all the samples, which corresponds to the $C_{it}$ level extracted experimentally by the C-V measurement. An additional $C_{it}$ peak with a Gaussian distribution at the peak energy of 0.1 eV below the CBE is introduced for sample 3, which corresponds to the observed hump.

## 2.2. $D_{it}$ extraction from the equivalent circuit analysis of C-V

The interface properties are studied via capacitance measurement for the same three samples. **Figure 3** shows corrected $C_{total}$-$V_{TG}$ curves for the frequency range of 1 kHz - 1 MHz. Parasitic capacitance ($C_{para}$) was carefully considered and removed (details are provided in **figure S4**). Ideally, the measured capacitance ($C_{total}$) is zero in the deep depletion region, that is, in the off state for the I-V, and saturates asymptotically to $C_{TG}$ in the strong accumulation region because $C_Q$ (~84 μFcm$^{-2}$) for monolayer MoS$_2$ in this region is much larger than $C_{TG}$. Therefore, all of the C-V curves at different frequencies were shifted to start from zero in the off state. This procedure is reasonable because the $C_{TG}$ obtained in the strong accumulation region after this correction is consistent with the $C_{TG}$ estimated from capacitive coupling between the top gate and back gate in the I-V within a 10% error. Hereafter, $C_{total}$ is defined as the measured capacitance without $C_{para}$. Frequency dispersion is observed in **figure 3** for all the samples. Observed frequency dispersion clearly indicates the interface quality of measured samples. Specifically, the sample with best interface quality (sample 1) has smallest frequency dispersion. In general, frequency dispersion has two origins. One comes from large $C_{it}$, which reveals the interface property directly [16]. The other comes from series resistance effect [17], even though the ohmic contacts are realized by the Ni/Au contact (Details are provided in **figure S5**). However, the large $C_{para}$ from SiO$_2$/$n^+$-Si substrate prevents us from extracting correct conductance signals. The quantitative analyses of series resistance effect is beyond the scope of this paper. The next task is to separately and quantitatively clarify $C_{it}$ and $C_Q$.

To quantitatively estimate the value of $D_{it}$ via its frequency response, the capacitance is measured as a function of frequency ($f$). The equivalent circuit is modeled as shown in **figure 4a**. $C_{total}$ can be calculated based on the following equation:

$$1/C_{total} = 1/C_{TG} + 1/(C_Q + C_{it})$$
$$= 1/C_{TG} + 1/[C_Q + e^2 D_{itA}(2\pi f \tau_{itA})^{-1} \arctan(2\pi f \tau_{itA}) + e^2 D_{itB}(2\pi f \tau_{itB})^{-1} \arctan(2\pi f \tau_{itB})], \quad (1)$$

where $\tau_{it}$ is the time constant for $D_{it}$, and A and B refer to two types of interface states. This equation is slightly different from that used in a previous paper for a CVD monolayer MoS$_2$ FET [16] because the multi-level model is more practical than the single-level model [20]. **Figure 4b** shows the $C_{total}$-$f$ curves at different $V_{TG}$ for sample 1. $C_{total}$ decreases with increasing frequency because $C_{it}$ is unable to completely respond at high frequency. Therefore, $1/C_{total}$ saturates to $1/C_{TG} + 1/C_Q$ at the high frequency limit according to eq. (1). By using $C_Q$ and two types of $D_{it}$ and $\tau_{it}$ as fitting parameters, the experimental data are well reproduced, as shown by the solid black lines. Although the number of fitting parameters is large, the accuracy of the estimated $D_{it}$ and $\tau_{it}$ is sufficiently high for quantitative analysis because $D_{it}$ and $\tau_{it}$ characterize different physical properties. Although two types of interface states are considered, $D_{it}$ mainly originates from the interface states of type A for most of the measured samples ($D_{itA} > D_{itB}$, $\tau_{itA} > \tau_{itB}$) (Details are provided in **figure S6**). Thus, $D_{itA}$ and $\tau_{itA}$ are simply referred to as $D_{it}$ and $\tau_{it}$. $C_Q$, $\tau_{it}$ and $D_{it}$ from this fitting are plotted as a function of $V_{TG}$ - $V_{TH}$ in **figure 4c,d**. $V_{TH}$ is theoretically defined as $V_{TG}$ at $C_Q = C_{TG}$ [19], which will be explained later in the $C_Q$ analysis. Both $C_Q$ and $\tau_{it}$ exhibit an approximately linear relation on the logarithmic scale[29,30]. It should be noted that $C_Q$ can only be accurately extracted when the saturation tendency of $C_{total}$ in **figure 4b** is clearly observed, which restricts the $V_{TG}$ range for the $C_Q$ extraction. Therefore, considering the $C_{total}$-$f$ curves for samples 2 and 3 shown in **figure S7**, $C_Q$ can be extracted from sample 2 but not from sample 3.

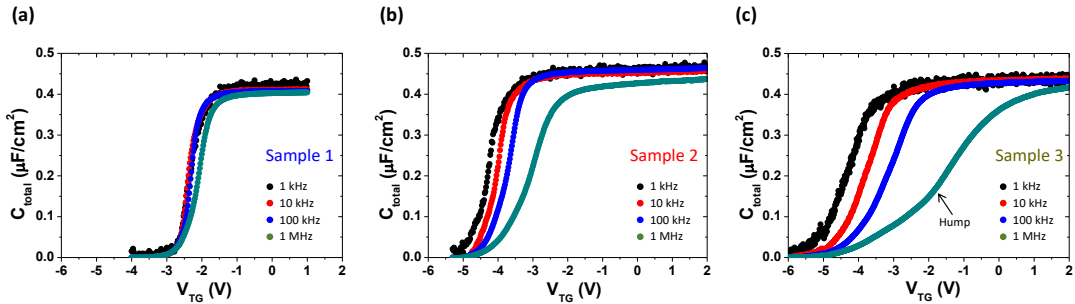

**Figure 3.** C-V characteristics of (a) sample 1, (b) sample 2, and (c) sample 3 with a frequency ranging from 1 kHz to 1 MHz. There is a clear correlation between the *S.S.* in the I-V characteristics and the degree of frequency dispersion.

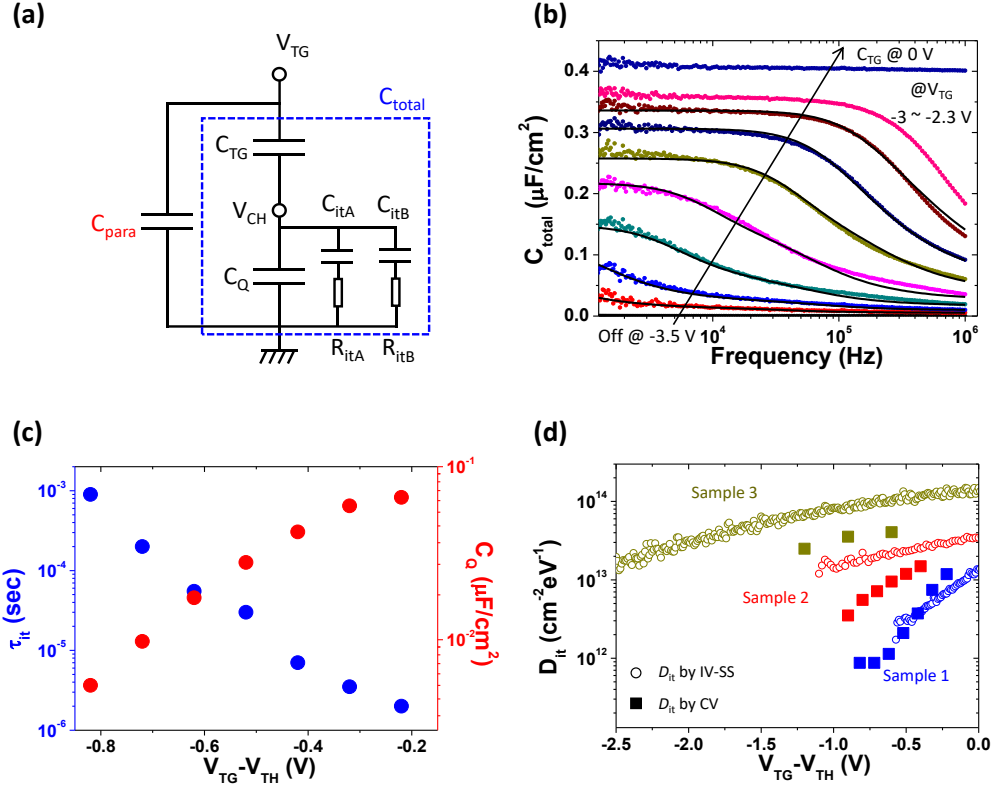

**Figure 4.** (a) The equivalent circuit model of the device. (b) $C_{total}$ as a function of frequency at different $V_{TG}$ (-3.0~-2.3 V with a step of 0.1 V) for sample 1. Solid circles are the experimental data, and black lines are the fitting curves generated via eq. (1). (c) Extracted $\tau_{it}$ and $C_Q$ as a function of $V_{TG}$-$V_{TH}$ for sample 1. (d) Extracted $D_{it}$ from both the C-V and I-V as a function of $V_{TG}$-$V_{TH}$ for the three samples.

Now, let us compare the $D_{it}$ values extracted from the C-V and I-V measurements. As shown in **figure 4d**, the values of $D_{it}$ are comparable. This indicates that the interface properties were successfully evaluated via the electrical measurements. The lowest $D_{it}$ obtained in this study is ~$8\times10^{11}$ cm$^{-2}$ eV$^{-1}$, which is one order of magnitude lower than that of a CVD MoS$_2$ FET [16]. It is expected that these improved interface properties come from a higher quality of the bulk MoS$_2$ crystals and the Y$_2$O$_3$ buffer layer. The $D_{it}$ tail close to the conduction band is still observed for all samples. It should be noted that $D_{it}$ is still presented as a function of $V_{TG}$ - $V_{TH}$, but not $E_F$, because the experimental correlation between $V_{TG}$ and the channel potential ($V_{CH}$, $E_F = eV_{CH}$) is not clear. In the next section, the effect of $C_{it}$ on $C_Q$ is discussed in detail, and as a result, the relation between $V_{TG}$ and $V_{CH}$ is revealed.

*2.3. Quantum capacitance analysis*

Quantum capacitance was originally derived from the finite *DOS* of a 2D electron gas [18,31,32]. In addition, it has been successfully extracted in graphene [33,34]. Here, using the same procedure as for graphene, which is different from the C-f analysis in **figure 4b**, $C_Q$ is again extracted as a continuous function of $C_{TG}$. The samples 1 and 2 are used for this analysis due to their relatively high qualities. First, $C_Q$ is extracted experimentally from the C-V measurements at 1 MHz in **figure 3** to observe the entire picture. At the high frequency limit of 1 MHz, the interface states are assumed to be unable to respond. Therefore, eq. (1) is reduced to $1/C_{total} = 1/C_{TG} + 1/C_Q$ by neglecting $C_{it}$. Since $C_{TG}$ has already been determined in the strong accumulation region, $C_Q$ for samples 1 and 2 is extracted experimentally as a function of $V_{TG}$ in **figure 5a,b**.

Alternatively, $C_Q$ can be calculated theoretically by considering the Fermi distribution and *DOS* of 2D materials and is expressed as follows [19]:

$$C_Q = e^2 g_{2D} \left[1 + \frac{\exp(E_G/2k_BT)}{2\cosh(eV_{CH}/k_BT)}\right], \qquad (2)$$

where $g_{2D} = g_s g_v m^*/2\pi\hbar^2$ is the band-edge *DOS*, and $E_G$ is 1.9 eV for monolayer $MoS_2$. $g_s$ and $g_v$ are the spin and valley degeneracy factors, respectively. $m^*$ is assumed to be $0.6m_0$, where $m_0$ is the electron mass in vacuum. The mid gap is defined to be $E_F = 0$ eV. Then, the conduction band edge (CBE) is located at 0.95 eV. In eq. (2), $C_Q$ is expressed as a function of $V_{CH}$, and the correlation between $V_{CH}$ and $V_{TG}$ is required for comparison with the experiment. Based on the ideal equivalent circuit at the high-frequency limit without $C_{it}$, the theoretical correlation between $V_{CH}$ and $V_{TG}$ is expressed as

$$V_{TG} = V_{TG,mid-gap} + \int_0^{V_{CH}} (C_Q + C_{TG})/C_{TG} dV_{CH}, \qquad (3)$$

where $V_{TG,mid-gap}$ is the fitting parameter that refers to $V_{TG}$ at $E_F = 0$ eV. This parameter is used to compensate for the intrinsic *n*-type doping in $MoS_2$. By combining eqs. (2) and (3), $C_Q$ is calculated as a function of $V_{TG}$. The experimental and theoretical $C_Q$-$V_{TG}$ curves are compared in **figure 5a**. The $C_Q$-$V_{TG}$ curve of sample 1 fits well with the theoretical curve over the wide range of $V_{TG}$ (-1.8 ~ 0.1 V), while it largely deviates from the theoretical curve for sample 2. However, even for sample 1, the deviation of $C_Q$ can be seen on the logarithmic scale, as shown in **figure 5b**.

The deviations of $C_Q$ from the theoretical curve along the transverse and vertical axes have two different origins. One, for the transverse axis, is the "stretch-out" effect [35]. Although the interface traps do not respond to the alternating current (AC) frequency of 1 MHz with the amplitude of 50 mV in the C-V measurement, they respond to the slowly varying direct current (DC) $V_{TG}$, which causes the C-V curve to stretch out along the transverse $V_{TG}$ axis as the interface trap occupancy changes with $V_{TG}$. The other origin, which impacts the vertical axis, is that the high-frequency limit of 1 MHz is not always satisfied since $\tau_{it}$ is quite short near the CBE, as shown in **figure 4c**. Thus, the extracted $C_Q$ may partially include the contribution of $C_{it}$ in terms of the vertical axis. As a result, the experimental correlation between $V_{CH}$ and $V_{TG}$ by a conventional high frequency method (the so-called Terman method) [20] needs to be reconsidered. For both cases, interface traps cause deviations from the theoretical $C_Q$ curve in the range of $C_Q < C_{it}$.

As discussed above, the stretch-out effect and the limitation of the measured frequency complicate the correlation between $V_{CH}$ and $V_{TG}$. Here, we propose a simple method to determine $V_{CH}$, i.e., find $E_F$ by using the $C_Q$ values obtained from the C-f analysis in **figure 4c** instead of the $C_Q$ values extracted from the C-V measurement at 1 MHz, because they do not include $C_{it}$. **Figure 5c** shows the theoretical $C_Q$-$E_F$ curve calculated via eq. (2). Experimental $C_Q$ values extracted from C-f analysis for samples 1 and 2 are then plotted on the theoretical $C_Q$-$E_F$ line as the blue open circles and red open triangles in **figure 5c**, which compensates the contribution of stretch out along the transverse axis in the C-V curve. Then, the correlated $E_F$ value can be read by following the arrows. We have to emphasize that $C_Q$ obtained via the *C-f* analysis is significantly more accurate than that obtained from the *C-V* curve at 1 MHz because $C_{it}$ can be strictly excluded from $C_Q$. Sample 2 has a narrower $E_F$ range due to its larger $D_{it}$. This means that modulation of $E_F$ by $V_{TG}$ is suppressed by a larger $D_{it}$, which is often referred to as Fermi level pinning at the semiconductor/insulator interface [36,37,38].

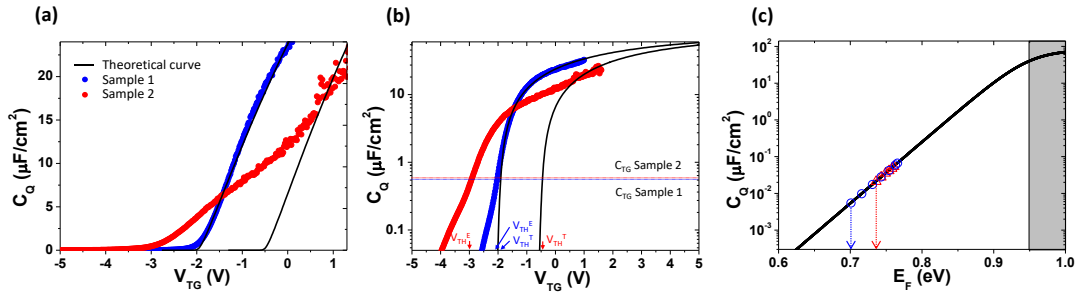

**Figure 5.** (a) Comparison of experiments and theory for $C_Q$ as a function of $V_{TG}$ for samples 1 and 2. $C_Q$ is experimentally extracted from the C-V curve at 1 MHz. The black line is the theoretical curve calculated based on eqs. (2) and (3). (b) The vertical axis of (a) is converted to the logarithmic scale. $V_{TH}^T$ and $V_{TH}^E$ indicate the theoretical and experimental $V_{TH}$ values, respectively, at $C_Q = C_{TG}$. (c) Theoretical $C_Q$-$E_F$ curve used for the correlation between $V_{TG}$ and $E_F$. Blue open circles and red open triangles indicate experimental $C_Q$ values extracted from the C-f analysis for samples 1 and 2, respectively.

The stretch-out effect due to the large $D_{it}$ can be clearly understood by comparing the $I_{DS}$-$V_{TG}$ curves from figure 2a and the $C_Q$-$V_{TG}$ curves from figure 5b. Theoretically, $V_{TH}$ is defined by $V_{TG}$ at $C_Q = C_{TG}$ in figure 5b [19]. The $V_{TH}$ determined experimentally for sample 2 is considerably shifted to the negative $V_{TG}$ direction due to stretch out by the large $D_{it}$. This situation is consistent with the $V_{TH}$ position in the $I_{DS}$-$V_{TG}$ curve, as shown in figure 2a. It is evident that the apparent $V_{TH}$ position in the *I-V* is largely affected by the degree of $D_{it}$.

### 2.4. Temperature-dependence of C-V & Physical origin for $D_{it}$

The slope of $C_Q$ becomes sharp at low temperatures due to the intrinsic nature of the Fermi distribution, which provides an alternative means to confirm the validity of $C_Q$ extraction. Based on this idea, both *C-V* and *I-V* measurements were performed at 75, 150 and 300 K for an additionally prepared monolayer $MoS_2$ FET that has a relatively high quality (two-probe mobility ~ 10 cm$^2$ V$^{-1}$ s$^{-1}$, S.S. = 240 mV/dec at room temperature).

$C_Q$ is again extracted from the *C-V* curves at 1 MHz and fitted as a function of $V_{TG}$ at different temperatures, as shown in **figure 6a,b**. The extracted $C_Q$-$V_{TG}$ curves are divided into two regions. The first region is the $C_Q$ dominant region, with $C_Q > C_{it}$. In this region, $C_Q$ has a clear temperature dependence and fits well with the theoretical calculation. The other region is the $C_{it}$ dominant region, with $C_Q < C_{it}$. The $C_Q$-$V_{TG}$ curve deviates from the theoretical curve and shows a gradual change with decreasing temperature. Here, let us consider the $V_{TH}$ shift with decreasing temperature in the $C_Q$-$V_{TG}$ relation, as shown in **figure 6b**. It is clear that $V_{TH}$ shifts positively with decreasing temperature due to the temperature dependences of $C_Q$ and $D_{it}$. This is quite important for studying temperature-dependent transport properties and is discussed later in relation to MIT.

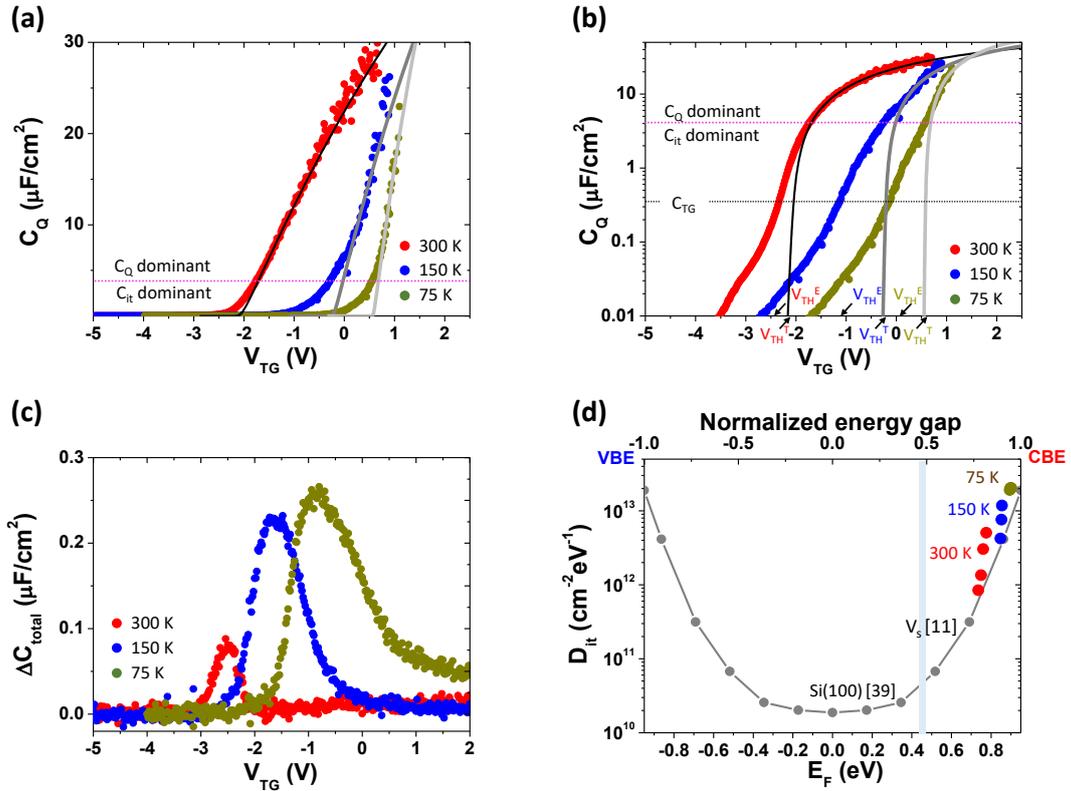

**Figure 6.** (a) $C_Q$ extracted as a function of $V_{TG}$ at 300 K, 150 K and 75 K for an additionally prepared monolayer $MoS_2$ FET. Solid lines are the theoretical fitting curves based on eqs. (2) and (3). (b) The same figure as (a) with the logarithmic scale. $V_{TH}^T$ and $V_{TH}^E$ indicate theoretical and experimental $V_{TH}$ values at $C_Q = C_{TG}$, respectively, for different temperatures. (c) $\Delta C_{total}$ calculated between 10 kHz and 1 MHz as a function of $V_{TG}$ at different temperatures. (d) $D_{it}$ as a function of $E_F$. VBE is the valance band edge. For comparison, $D_{it}$ for Si(100) [39] is also included as a function of normalized energy gap (top transverse axis).

To show the frequency dispersion at different temperatures, the capacitance difference ($\Delta C_{total}$) between 10 kHz and 1 MHz as a function of $V_{TG}$ at different temperatures is shown in **figure 6c**. $\Delta C_{total}$ gradually increases and broadens with decreasing temperature, which is reasonable for the band tail behavior. The exact $D_{it}$ value is extracted based on eq. (1). Having confirmed the $C_Q$ analysis at the measured temperatures, $D_{it}$ is illustrated as a function of $E_F$, as shown in **figure 6d**. In addition, the band tail distribution of $D_{it}$ is successfully confirmed using the temperature-dependent C-V measurements, the results of which are similar to those of the $SiO_2$/Si case [39].

Let us discuss the physical origin of $D_{it}$ for a monolayer of $MoS_2$. According to the DFT calculation [11], an S vacancy introduces an isolated $D_{it}$ peak at 0.46 eV below the CBE, which is also indicated in **figure 6d**. It is clear that the present band tail behavior of $D_{it}$ is not directly related to the S vacancy. This band tail distribution of $D_{it}$, which is also called the U-shaped band edge states, has been widely observed in $Si/SiO_2$ [39-41], $Ge/GeO_2$ [42,43] and other conventional oxide semiconductor interfaces [44]. In the case of the $Si/SiO_2$ interface, many models have been proposed to explain the U-shaped band edge states. For example, the stretched Si-Si bonds at the interface [40] and distortion of the Si-O-Si bond angle [45] are expected to cause trap levels because the conduction band is composed of an anti-bonding state of the $sp^3$ hybrid orbital. The strain is concentrated at the $Si/SiO_2$ interface due to the density difference. This may cause a deviation of the anti-bonding state energy, resulting in the U-shaped band edge states. Therefore, since the conduction and valence bands of $MoS_2$ are mainly composed of the energy splitting of the Mo $d$ orbital [11, 46], the Mo-S bond bending due to the strain caused by lattice mismatch at the $MoS_2$/high-$k$ interface [47,48], the surface roughness of the $SiO_2$ surface, or bond bending related to the S vacancy might be the origin. However, further study is required to clarify the physical origin of the U-shaped $D_{it}$ in $MoS_2$.

The interface properties of a bulk $MoS_2$ capacitor have been measured as an isolated $D_{it}$ peak by using the Terman method [17] and it is suggested that it be ascribed to the S vacancy. Our multilayer $MoS_2$ FET also shows the hump in C-V curves (data is not shown here). More interestingly, a similar hump is also observed in monolayer $MoS_2$ with poor interface quality (sample 3), as shown by arrows in the C-V curve of **figure 3c** and the $I_{DS}$-$V_{TG}$ curve of **figure 2b**. Thus, the origin of this hump could be the sulfur vacancy or its derivative (e.g., disulfur vacancy). However, for samples 1 and 2 with relatively high quality interfaces, the hump is not obvious. Since the $D_{it}$ level of the band tail of a monolayer is much higher than that of a multilayer, the high $D_{it}$ level of a monolayer may hide the isolated $D_{it}$ peak of the S vacancy. If this is the case, the observation of the hump in sample 3 suggests that the concentration of S vacancies in sample 3 is highest.

Let us discuss the reason why the C-V curves of a monolayer $MoS_2$ have a significantly larger frequency dispersion than those of Si even though the $D_{it}$ energy distributions are roughly comparable. One reason is that the large band-gap of the monolayer $MoS_2$ broadens the $D_{it}$ energy distribution. The most important reason is the smaller DOS of the monolayer $MoS_2$. As we discussed in the previous section, the effect of $C_{it}$ on $C_{total}$ is determined by the relative ratio of $C_{it}/C_Q$. When $C_Q$ is smaller than $C_{it}$ over a certain energy range, $C_{it}$ degrades the C-V curve in terms of the large frequency dispersion, large $V_{TH}$ shift, limited modulation of $E_F$ by $V_{TG}$, and other factors. In the case of the Si FET structure, semiconductor capacitance is composed of inversion layer capacitance ($C_{Inv}$) instead of $C_Q$, as shown in **figure S1**. The DOS for Si inversion is much larger than that for the monolayer $MoS_2$, which suppresses the effect of $C_{it}$ on $C_{total}$. This is supported by the reduced frequency dispersion in the C-V curve for the multilayer $MoS_2$ due to the larger $C_Q$. As a result, ultra-thin 2D materials are more sensitive to interface disorder due to reduced DOS.

## 2.5. $C_Q$ and $C_{it}$ effect on I-V characteristics

Since all the constituents in $C_{total}$ are well understood, it is now possible to reproduce I-V characteristics by completing two steps: the determination of the carrier density controlled by the electrostatic field-effect of the top gate, and then, the characterization of the electron transport of these carriers in the conduction band. Therefore, carrier density control by $V_{TG}$ is modeled based on the well-understood equivalent circuit. $V_{CH}$ can be correlated to $V_{TG}$ as follows:

$$V_{TG} = V_{TG,mid-gap} + \int_0^{V_{CH}} (C_Q + C_{it} + C_{TG})/C_{TG} dV_{CH} . \tag{4}$$

For this equation, $C_{it}$ is added to eq. (3) because $C_{it}$ is able to respond completely in conventional I-V characteristics due to the DC measurement. The channel carrier density ($n_{CH}$) - $V_{CH}$ relation is calculated using the equation [19]

$$n_{CH} = g_{2D} k_B T \ln\left[1 + \exp\left((eV_{CH} - E_{CBE})/k_B T\right)\right]. \tag{5}$$

Then, the fundamental drift current equation $\sigma = en_{CH}\mu_D$ is applied to simulate the carrier transport process, where $\sigma$ and $\mu_D$ refer to conductivity and drift mobility, respectively. Since $C_Q(E_F)$ is analytically calculated, $C_{it}(E_F)$, $\mu_D(E_F)$ and $V_{TG,mid-gap}$ are used as fitting parameters. $\mu_D(E_F)$ is assumed to be independent of $E_F$ with a constant value for simplicity. Then, the drift mobilities for samples 1, 2 and 3 are estimated to be 12.3, 8.2, 2.3 $cm^2$ $V^{-1}$ $s^{-1}$, respectively.

Drift mobilities are slightly higher than two-probe field-effect mobilities obtained experimentally (**Table S1**) because $C_{it}$ reduces carrier controllability by the gate even in the linear region of $I_{DS}$-$V_{TG}$ curves. Although this is a rough assumption, it is valid in the linear region of the *I-V*. Whereas in the subthreshold region, the drift mobility might decrease with reduced screening effect, the dominant factor in determining $I_{DS}$ in this region is the carrier density, which is exponentially related to $V_{TG}$, instead of the drift mobility. Band-tail-shaped $C_{it}$ curves with three different levels are assumed for the samples, as shown in **figure 2c**. For sample 3, an additional $C_{it}$ peak with a Gaussian distribution at the peak energy of 0.1 eV below the CBE is introduced, which is used to reproduce the hump observed in **figure 2b**. Although the peak energy for a single S vacancy is reported to be 0.46 eV below the CBE by DFT calculation [11], the energy level of the present $C_{it}$ peak is quite shallow, suggesting the existence of clustering of S vacancies [49]. Finally, experimental $I_{DS}$-$V_{TG}$ characteristics of samples 1, 2 and 3 are well fitted based on the above model, as shown by the black solid lines in **figure 2a,b**. Additionally, an ideal *I-V* curve without $C_{it}$ for sample 1 is exhibited by the green solid line in **figure 2a,b**, where the ideal *S.S.* of ~60 mV/dec as well as the sharp transition from the linear to the subthreshold region are evident. In this case, $V_{TH}$ can be uniquely determined. However, $C_{it}$ does degrade the subthreshold region, i.e., a *S.S.* of ~100 mV/dec for sample 1 and over 300 mV/dec for samples 2 and 3. This degradation introduces the ambiguity in the $V_{TH}$ extraction by experiment, which has been encountered in the *C-V* analysis as well.

## 2.6. Interpretation of MIT

In the final section, let us discuss the contribution of $C_{it}$ to MIT. The top gate FET structure in **figure 1b** is unsuitable for studying $I_{DS}$-$V_G$ in the linear region precisely due to the existence of the access region, which results in the underestimation of the intrinsic drift mobility in the linear region. Thus, a back-gate four-probe FET with monolayer MoS$_2$ on 90-nm SiO$_2$/$n^+$-Si substrate is prepared. The experimental $\sigma$-$V_{BG}$ curves excluding the series resistance are shown in **figure 7a**. Clear MIT behavior is observed for the present device quality. So far, MIT of MoS$_2$ has been discussed for both *I-V* [8,9] and *C-V* [10] with different models. Here, the temperature dependences of $\sigma$-$V_{BG}$ curves obtained experimentally are again reproduced by the above-mentioned model using the relation $\sigma = en_{CH}\mu_D$. The temperature dependence of $C_Q$ is calculated in **figure S8a**, which results from the natural property of the Fermi distribution. The band tail shape of $C_{it}$ with three different levels, that is, high, low and no $C_{it}$, is again assumed in **figure S8a**, while $C_{BG}$ is estimated as 0.038 µFcm$^{-2}$ for back gate SiO$_2$ oxide with a thickness of 90 nm. $V_{BG,mid-gap}$, instead of $V_{TG,mid-gap}$, is constant for all temperatures. Then, $n_{CH}$ can be calculated using eqs. (4) and (5). Moreover, the $\mu_D$ used for this modeling is the same for all three $C_{it}$ cases and is slightly larger than the experimental $\mu_{FE}$ at all temperatures (Details are shown in **figure S8b**). **Figure 7b** shows simulated $\sigma$-$V_{BG}$ curves with three different $C_{it}$ levels. MIT is well reproduced using the high $C_{it}$. By decreasing the $C_{it}$ level, the crossover points of the MIT shift to the negative $V_{BG}$ side and finally enter the subthreshold region for the case with no $C_{it}$, which blinds the MIT. Recently, no MIT has been reported for an *h*-BN-encapsulated monolayer CVD-MoS$_2$ FET [50], suggesting a quite low $C_{it}$ due to superior 2D/2D interface properties.

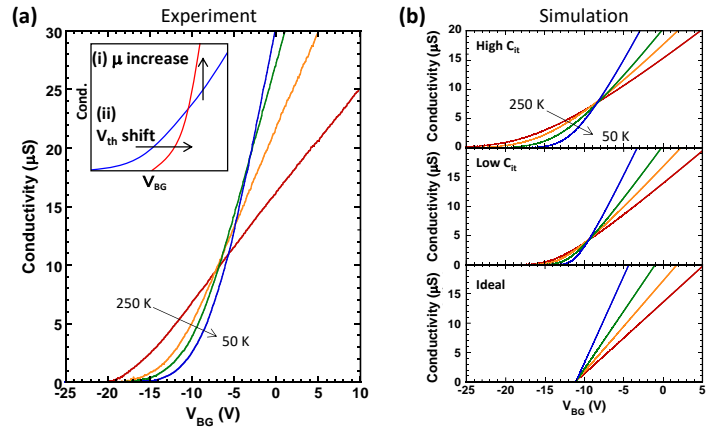

**Figure 7.** (a) Experimental four-probe conductivity as a function of $V_{BG}$ for a back-gate monolayer MoS$_2$ FET at 50~250 K, showing the typical MIT behavior. (b) Simulated $\sigma$-$V_{BG}$ curves with high $C_{it}$, low $C_{it}$ and no $C_{it}$ (ideal), which are shown in figure S8a. $V_{BG,mid-gap}$ =-53 V, -23 V, and -12 V are assumed for high $C_{it}$, low $C_{it}$ and no $C_{it}$, respectively.

Generally, MIT can be observed intuitively by the combination of (i) the increase in the mobility and (ii) positive $V_{TH}$ shift with decreasing the temperature. Within the present model, the mobility is assumed to increase with decreasing temperature due to suppression of phonon scattering, as observed in the experiment. Therefore, the dominant key factor for MIT is a positive $V_{TH}$ shift with decreasing temperature. This occurs because $E_F$ at $V_{TH}$ approaches the CBE at lower temperature. Thus, a larger amount of $C_{it}$ needs to be filled by electrons before reaching $V_{TH}$ at lower temperature. This also explains why $V_{TH}$ shifts more with temperature in the high $C_{it}$ case. So far, many models have been developed for MIT on 2D layered channels. The present model indicates that $C_{it}$-induced positive $V_{TH}$ shift is one of the main origins for "extrinsic" MIT.

## 3. Conclusion

The degradation of the electrostatic field-effect control for the monolayer mechanical exfoliated MoS$_2$ FET is systematically studied using both C-V and I-V characterization in terms of $C_Q$ and $C_{it}$. $C_Q$ was confirmed over all of the measured temperature ranges (75~300 K). Therefore, $D_{it}$ was evaluated as a function of $E_F$ by the newly constructed $C_Q$ analysis, which can also be applied for other monolayer TMDs. $D_{it}$ was extracted as $10^{12}$~$10^{13}$ cm$^{-2}$ eV$^{-1}$ with a band tail shape close to the conduction band, which is comparable to that in Si/SiO$_2$. However, ultra-thin 2D materials are more sensitive to interface disorder due to the reduced DOS, which drastically degrades the subthreshold properties. The multilayer MoS$_2$ is more suitable for device application due to its larger DOS. Having elucidated all the constituents in $C_{total}$ quantitatively by C-V measurements, I-V characteristics are then well reproduced and understood by utilizing the drift current model. One of the physical origins for MIT is suggested to be the extrinsic outcome of the $V_{TH}$ shift due to $C_{it}$ and $C_Q$. Capacitance measurement is quite informative for detecting interface states and density of states in ultra-thin 2D materials, which allows us to understand device physics and improve device performance.

**Acknowledgements**
N. Fang was supported by a Grant-in-Aid for JSPS Research Fellows from the JSPS KAKENHI. This research was supported by the JSPS Core-to-Core Program, A. Advanced Research Networks, JSPS KAKENHI Grant Numbers JP25107004, JP16H04343, JP16K14446, and JP26886003, and JST PRESTO Grant Number JPMJPR1425, Japan.

Supplementary data

# Band tail interface states and quantum capacitance in a monolayer molybdenum disulfide field-effect-transistor

**Nan Fang[1] and Kosuke Nagashio[1,2]**

[1]Department of Materials Engineering, The University of Tokyo, Tokyo 113-8656, Japan

[2]PRESTO, Japan Science and Technology Agency (JST), Japan

E-mail: nan@adam.t.u-tokyo.ac.jp, nagashio@material.t.u-tokyo.ac.jp

# Supplementary data

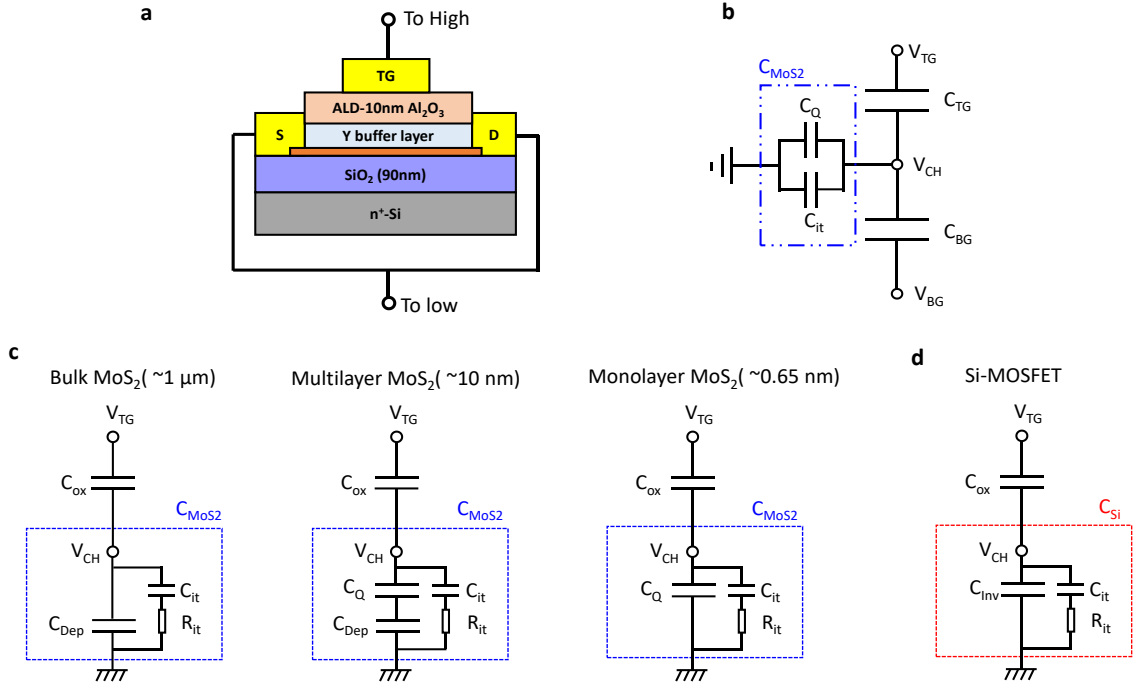

**Figure S1.** (a) The configuration of capacitance measurement for MoS$_2$ FET. (b) Simplified equivalent circuit of monolayer MoS$_2$ in dual-gate *I-V* measurement. (c) Three different equivalent circuits for the MoS$_2$ FET with different channel thickness for *C-V* measurement. $C_{Dep}$ is expressed as $\varepsilon_S/W_D$, where $\varepsilon_S$ is dielectric constant of semiconductor and $W_D$ is depletion width. It should be noted that MoS$_2$ FET operates at the accumulation region and turns off at the depletion region. Therefore, for all three cases, the inversion layer capacitance ($C_{Inv}$) is not shown. (d) Si MOSFET operates at the inversion region. Therefore, for the *C-V* measurement of Si-MOSFET structure, the semiconductor capacitance is composed of $C_{Inv}$, instead of $C_Q$.

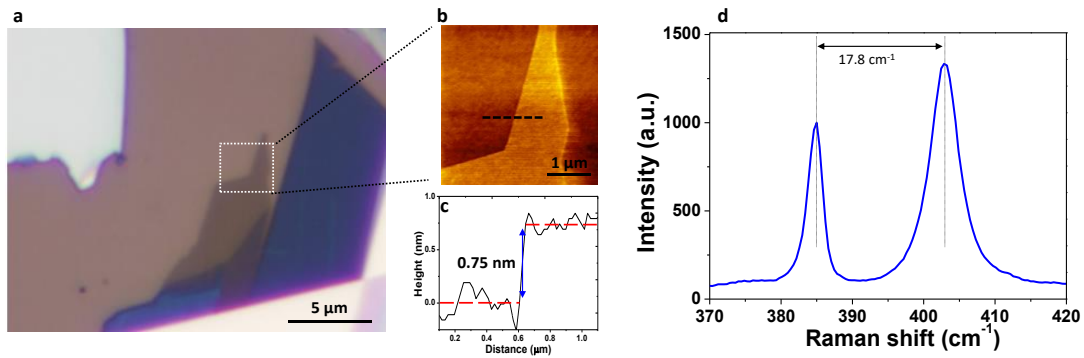

**Figure S2.** Thickness identification of MoS$_2$. (a) Optical image of MoS$_2$ with different layer numbers. Monolayer can be simply identified in this image by the contrast, which is then confirmed by AFM and Raman spectroscopy. (b) AFM image of the white-dashed rectangle area in (a). (c) Height profile along the dashed line in (b). The height of monolayer MoS$_2$ is 0.75 nm. (d) Raman spectroscopy of typical measured monolayer MoS$_2$. The wavenumber between the two peaks (E$_{2g}$, A$_{1g}$) is 17.8 cm$^{-1}$, which proves to be monolayer MoS$_2$.

# Supplementary data

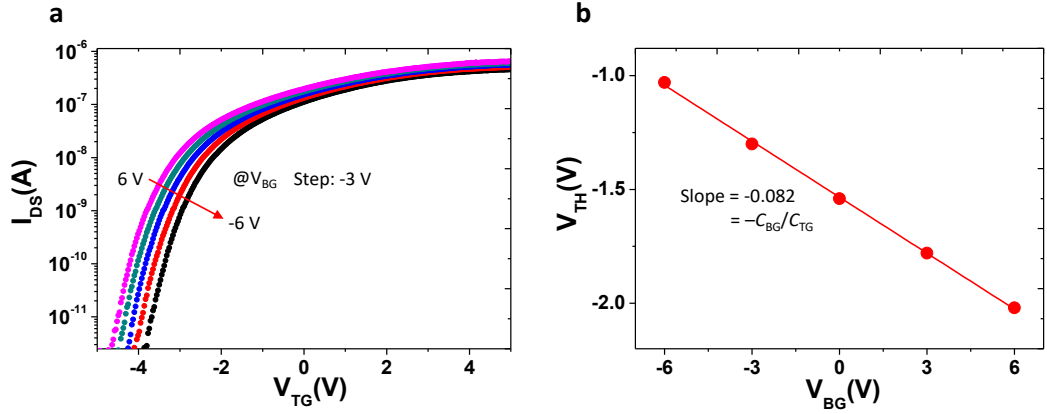

**Figure S3.** (a) $I_{DS}$–$V_{TG}$ characteristics for sample 3 with $V_{DS} = 0.1$ V at different $V_{BG}$. $V_{BG}$ ranges from 6~-6 V with a step of 3 V. (b) The trace of $V_{TH}$ observed for the $V_{TG}$ sweep as a function of $V_{BG}$. The $V_{TH}$ position is controlled by the relative ratio of capacitive coupling between the top and back gates with a MoS$_2$ channel (Detailed derivation are shown below). Therefore, the slope shown in (b) corresponds to $-C_{BG}/C_{TG}$, where $C_{BG}$ and $C_{TG}$ are the back and top gate capacitances, respectively. Because $C_{BG}$ is 0.038 µF/cm² for the 90-nm SiO$_2$ with $k_{SiO2} = 3.9$, $C_{TG}$ can be estimated to be 0.46 µF/cm².

Based on equivalent circuit in figure S1b, the total charge in the channel is induced by both top gate and bottom gate.

$$V_{CH}C_{MoS2} = (V_{TG} - V_{CH})C_{TG} + (V_{BG} - V_{CH})C_{BG}, \qquad (1)$$

where $C_{MoS2}$ is capacitance of MoS$_2$, which consists of $C_Q$ and $C_{it}$ in parallel. Carrier density is constant in principle when source/drain current is kept unchanged. As a result, $V_{CH}$ and $C_{MoS2}$ are also constant[1]. Therefore, after transformation,

$$Const. = V_{CH}(C_{MoS2} + C_{TG} + C_{BG}) = V_{TG}C_{TG} + V_{BG}C_{BG}. \qquad (2)$$

By modulating top gate and bottom gate simultaneously at constant source/drain current, $V_{BG} \rightarrow V_{BG} + \Delta V_{BG}$, $V_{TG} \rightarrow V_{TG} + \Delta V_{TG}$.

Equation 2 turns to be

$$Const. = (V_{TG} + \Delta V_{TG})C_{TG} + (V_{BG} + \Delta V_{BG})C_{BG}. \qquad (3)$$

By comparing equation 2 and 3,

$$\Delta V_{TG}C_{TG} + \Delta V_{BG}C_{BG} = 0, \qquad (4)$$

or

$$-C_{BG}/C_{TG} = \Delta V_{TG}/\Delta V_{BG}. \qquad (5)$$

Experimentally, source/drain current is kept as constant below or close to the current level at $V_{TH}$ due to high sensitivity of carrier density as a function of gate bias at subthreshold region.

# Supplementary data

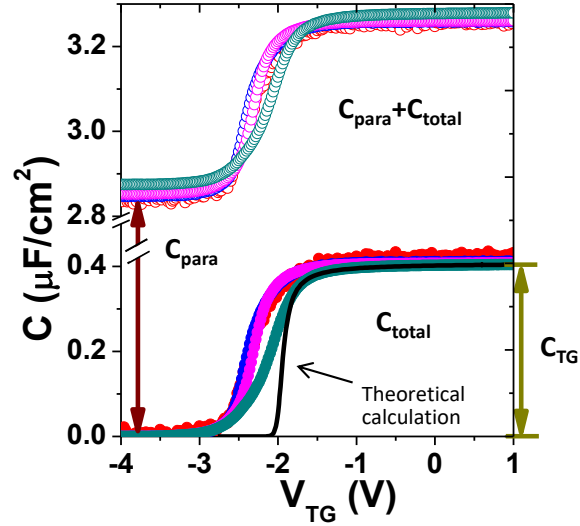

**Figure S4.** The open circles indicate the measured raw *C-V* data at different frequencies for sample 1, which includes $C_{para}$. It should be noted that the back gate voltage was not applied, just floating. Ideally, the measured capacitance ($C_{total}$) is zero at the deep depletion region, that is, the off state for *I-V*, and saturates asymptotically to $C_{TG}$ at the strong accumulation region because $C_Q$ (~84 µF cm$^{-2}$) for monolayer MoS$_2$ at this region is much larger than $C_{TG}$. Therefore, all of the *C-V* curves at different frequencies were shifted to start from zero at the off state. The solid circles indicate $C_{total}$ after removing $C_{para}$. After this correction, $C_{total}$ saturate asymptotically to $C_{TG}$ at the strong accumulation region. Therefore, $C_{TG}$ is extracted at accumulation region, which is consistent with $C_{TG}$ estimated from *I-V* in figure S2 within the 10% error. Reversely, the black line is theoretical calculation based on $1/C_{total}=1/C_Q + 1/C_{TG}$, where $C_Q$ was calculated using eq. (3) in the main text and the constant $C_{TG}$ value obtained above was used in this calculation. At the saturation region, theoretical calculation successfully reproduced the *C-V* data, suggesting that the present correction is reasonable.

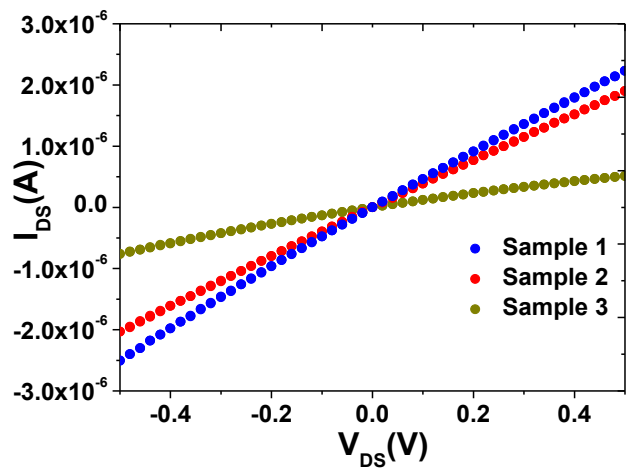

**Figure S5.** $I_{DS}$-$V_{DS}$ curve without gate bias at 300 K. The linear properties indicate good ohmic contact for all three samples.

# Supplementary data

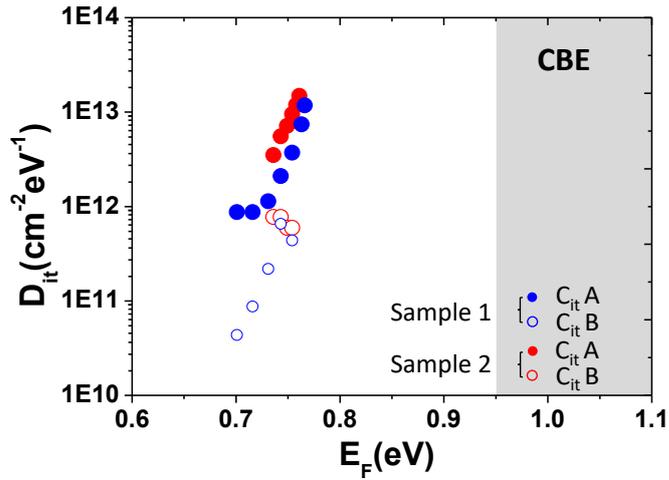

**Figure S6.** $D_{it}$ is shown as a function of $E_F$ by referring to $E_F$-$C_Q$ correlation, which is discussed in the Quantum capacitance analysis section. $D_{itA}$ is the main interface states which shows tail distribution close to conduction band edge and dominate interface properties. $D_{itB}$ is much smaller than $D_{itA}$ and not always observed. Thus, $D_{itB}$ is not discussed for samples 1, 2 in the main text.

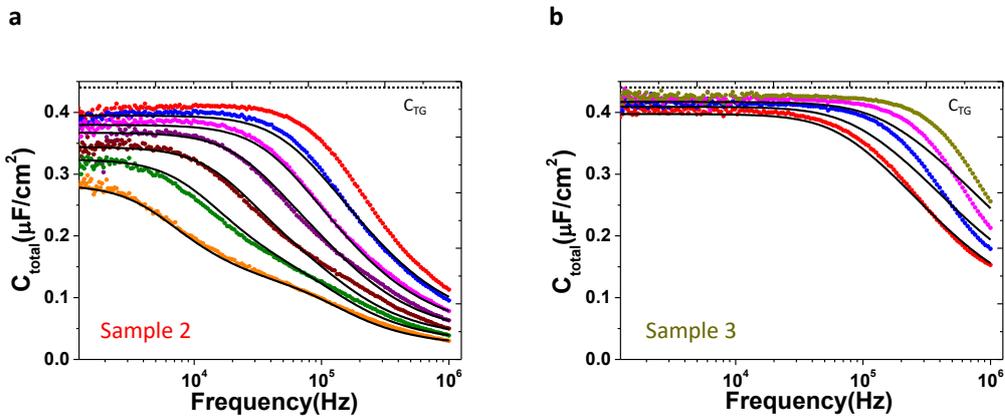

**Figure S7.** (a) Capacitance as a function of frequency for sample 2 at fixed $V_{TG}$ (-3.8 ~ -3.2 V with the step of 0.1 V). (b) Capacitance as a function of frequency for sample 3 at fixed $V_{TG}$ (-2.4 ~ -1.5 V with the step of 0.3 V). Solid circles are experimental results and black solid lines are fitting by using eq. (1) in the main text.

# Supplementary data

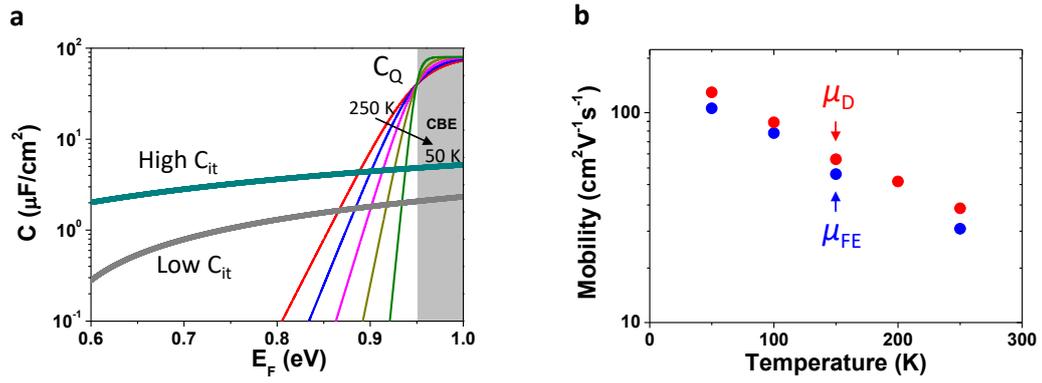

**Figure S8.** (a) $C_{it}$ of with different levels and calculated $C_Q$ as a function of $E_F$ at different temperatures (50 ~ 250 K) used for *I-V* modelling based on back-gate monolayer MoS$_2$ FET on 90-nm SiO$_2$/$n^+$-Si substrate. Ideally, $C_{it}$ does not show any temperature dependence, while $C_Q$ shows clear temperature dependence from Fermi distribution. It should be noted that the temperature dependence of $C_Q$ shown in figure 6a,b in the main text is shown as a function of $V_{TG}$. (b) Comparison of $\mu_{FE}$ estimated from the experimental *I-V* data and $\mu_D$ used as fitting parameter in the modelling.

# Supplementary data

**Note S1.** $D_{it}$ **extraction from S.S. in** *I-V*.

The definition of S.S at subthreshold region is[2]

$$S.S. = (\ln 10) \frac{dV_G}{d(\ln I_D)} = (\ln 10) \frac{k_B T}{e} \frac{dV_G}{d(V_{CH})}. \tag{6}$$

Based on equivalent circuit and considering the fact that $C_Q$ at subthreshold region is much smaller than $C_{ox}$ and $C_{it}$,

$$\frac{d(V_G)}{d(V_{CH})} = \frac{C_{TG} + C_Q + C_{it}}{C_{TG}} \approx \frac{C_{TG} + C_{it}}{C_{TG}}. \tag{7}$$

Since S.S. is related with $V_{TG}$ experimentally,

$$S.S.(V_{TG}) = (\ln 10) \frac{k_B T}{e} \frac{C_{TG} + C_{it}(V_{TG})}{C_{TG}}, \tag{8}$$

$D_{it}$ is then extracted as a function of $V_{TG}$.

**Table S1. Physical properties extracted from measured devices.**

|  | Monolayer Sample 1 | Monolayer Sample 2 | Monolayer Sample 3 | Monolayer T-dependence | Multilayer ~9 nm |
|---|---|---|---|---|---|
| Gate area (μm²) | 143 | 105 | 133 | 37 | 418 |
| $C_{ox}$ by IV (μF/cm²) | 0.38 |  | 0.46 | 0.34 |  |
| $C_{ox}$ by CV (μF/cm²) | 0.41 | 0.44 | 0.44 | 0.36 | 0.31 |
| Mobility at 300 K (cm²V⁻¹s⁻¹) | 9.5 | 6.0 | 2.0 | 10.0 | 33.0 |
| S.S. at 300 K (mV/dec) | 110 | 300 | 300 | 240 | 160 |

$C_{TG}$ values for three samples are different because they are from different batches.